# Transferable and Robust Machine Learning Model for Predicting Stability of Si Anodes for Multivalent Cation Batteries

Joy Datta, Dibakar Datta, Vidushi Sharma*

Department of Mechanical and Industrial Engineering, New Jersey Institute of Technology,

Newark, New Jersey 07103, United States

*Corresponding author: Vidushi Sharma, Email: vs574@njit.edu, vidushis@ibm.com

## Abstract

Data-driven methodology has become a key tool in computationally predicting material properties. Currently, these techniques are priced high due to computational requirements for generating sufficient training data for high-precision machine learning models. In this study, we present a Support Vector Regression (SVR)-based machine learning model to predict the stability of silicon (Si) – alkaline metal alloys, with a strong emphasis on the transferability of the model to new silicon alloys with different electronic configurations and structures. We elaborate on the role of the structural descriptor in imparting transferability to the model that is trained on limited data (~750 Si alloys) derived from the Material Project database. Three popular descriptors, namely X-Ray Diffraction (XRD), Sine Coulomb Matrix (SCM), and Orbital Field Matrix (OFM), are evaluated for representing Si alloys. The material structures are represented by descriptors in the SVR model, coupled with hyperparameter tuning techniques like Grid Search CV and Bayesian Optimization (BO), to find the best performing model for predicting total energy, formation energy and packing fraction of the Si alloy systems. The models are trained on Si alloys with lithium (Li), sodium (Na), potassium (K), magnesium (Mg), calcium (Ca), and aluminum (Al) metals, where Si-Na and Si-Al systems are used as test structures. Our results show that XRD, an experimentally derived characterization of structures, performs most reliably as a descriptor for total energy prediction of new Si alloys. The study demonstrates that by qualitatively selection of training data, using hyperparameter tuning methods, and employing appropriate structural descriptors, the data requirements for robust and accurate ML models can be reduced.

# 1. Introduction

Increasing demand for electric vehicles (EVs) has highlighted the energy storage limitations of commercial graphite anode-based lithium-ion batteries (LIBs). Energy storage in graphite with an intercalation mechanism offers low gravimetric energy densities of 372 mAhg$^{-1}$[1]. Alternatively, energy storage in electrodes through a conversion mechanism can promise a tenfold improvement in energy densities. The most popular anode after graphite is Silicon (Si), which has a gravimetric energy density of 3572 mAhg$^{-1}$ [2]. Si reacts with incoming Li to form an alloying mixture of Li$_x$Si during battery charging [3]. Furthermore, there is a pressing issue surrounding the scarcity of Li for LIBs. To meet the requirements of the future EV industry, we cannot rely solely on non-renewable Li[4]. Active research efforts are being made to develop advanced battery technologies beyond Li ion[5]. Due to the high capacity offered by Si-Li anode, Si has found applications in alkali earth metal batteries such as sodium (Na) ion batteries[6], magnesium (Mg) ion batteries[7], and

calcium(Ca) ion batteries[8] , to name a few. Similar to Si-Li system, Si anode reacts with Mg to form $Mg_2Si$ phase with a gravimetric density of 3816 mAhg$^{-1}$ [9], and reacts with Ca to form $Ca_2Si$ alloys with a maximum theoretical capacity of 3818 mAhg$^{-1}$ [8]. However, Si anode face challenges related to structural stability and volume expansion (~ 300% for LIBs) that lead to premature fractures, capacity losses, and limited cycle life of batteries[10–12]. Therefore, before designing and experimenting such battery materials, it is imperative to study the stability and structural assessment of Si-metal anodes computationally.

Over the last two decades, numerous computational efforts have been dedicated to understanding the Si-Li microstructures in the alloy mixtures and structural integrity of Si-based anodes [13–16]. Fan et al.[17] studied the effect of increasing Li concentration on the electro-mechanical stability of amorphous $Li_xSi$ anode by molecular dynamics (MD) simulations. They reported atomic bonding transitions from covalent to metallic bonds of a-$Li_xSi$ under various loading conditions. The study details how mechanical property of a-$Li_xSi$ changes during lithiation under different loading conditions. Similar atomistic simulation studies have also been conducted on Si anodes in alkali ion batteries[18,19]. It is evident that Si anodes possess greater potential than their competitor anodes for future multivalent cation batteries and will remain a subject of extensive research in the years to come[20,21].

Considering the stability of Si-based anodes during electrochemical cycling as a primary concern, it is necessary to study how Si anodes can maintain both low volume expansion and high capacity simultaneously. Exploring these possibilities of Si alloy anodes can be achieved by varying the compositions and stoichiometry ratio of Si and alloying multivalent cations. Prior to conducting experiments, it is essential to assess the synthesizability of any unknown structure based on its stability. Density functional theory (DFT) can be employed to predict the structural stability of materials using energy hull diagrams. This method involves calculating the formation energy of all possible stoichiometric ratios for a given composition using DFT. The structures with the lowest formation energy are considered the most stable, and an energy hull diagram can be created to illustrate the relative stability of different compositions. This approach has been successfully demonstrated in various studies [22,23]. Another important characteristic of a structure is the packing fraction, which describes the degree of porosity within the material. This parameter is expressed as a dimensionless quantity and represents the ratio of the total volume occupied by the crystal atoms to the volume of the unit cell. A low packing fraction indicates highest stability of an unknown material[24].

Theoretical studies based on simulations have proven to be valuable in providing design insights and performance predictions for experimental design[25]. Lately, the efficiency of computational simulations has diminished with the increasing complexity of materials. Quantum mechanics-based DFT methods are limited to small atomic systems consisting of approximately 200 atoms or fewer due to their computational expense[26]. On the other hand, classical Newtonian simulations like MD require less computational power and can be a viable alternative to simulating larger atomic systems. However, these techniques significantly abate the thermodynamic accuracy of disordered structures[27]. Furthermore, there is a lack of available interatomic potentials in the literature for newer material combinations [28].

Machine learning (ML) techniques have been widely adopted in material modeling to predict energy and simulate materials more efficiently[29–32]. ML is actively being researched to tackle various material science challenges, including property prediction, material discovery, and system optimization[33–35]. Common ML models used in this field include regression[36,37], gaussian approximation[38,39], graph neural networks (GNN)[40], and high-dimensional neural networks (HDNN)[41]. ML approaches in material systems can be categorized into graph-based and descriptor-based methods[42]. However, graph-based models may not be well-suited to handle long-range inter-atomic interactions. On the other hand, the descriptor-based approach offers more flexibility for feature engineering, allowing researchers to incorporate a broader range of atomic interactions in both spatial and dimensional aspects. Jihang et al.[43] predicted molecular property by four descriptor-based ML models and four graph-based models. The results revealed that the graph-based models required higher computational costs and resources compared to the descriptor-based models.

The selection of appropriate descriptors is crucial and should be based on the specific task at hand and their compatibility with the targeted machine learning algorithm[44–46]. Descriptors are numeric vectors that characterize the atomic or molecular structure and serve as inputs to ML models[42]. Various descriptors have been proposed for materials, including the coulomb matrix (CM)[47], sine coulomb matrix (SCM)[48], atom-centered symmetry functions (ACSF)[49], smooth overlap of atomic orbitals (SOAP)[50], and orbital field matrix (OFM)[51] , among others. Each descriptor has been designed to meet material science field but may not have broad applicability across different domains. By combining a well-defined descriptor and an appropriate model, AI techniques can leverage the wealth of experimental and simulation data available in established databases such as Material Project Database (MPD)[52], Open Quantum Materials Database[53], AFLOW[54] and Inorganic Crystal Structure Database(ICSD)[55], to address some of the most pressing predictive and discovery challenges among materials.

In this study, we predict the stability and packing fraction(PF) of Si alloys using the Support Vector Regression (SVR)[56] - based ML model. SVR is a supervised learning approach well-suited for handling non-linear regression problems[57]. It serves as an efficient alternative to more advanced methods like HDNN, which require extensive computational resources and data for training the model[58–60]. By coupling SVR with advanced statistical techniques, solid-state material properties can be predicted with high accuracy and minimal data requirements[61]. Thus, SVR is an ideal approach for rapidly predicting the stability of materials such as Si alloys, which have broad applications in the energy domain[62–66] but limited available data. We utilize the Si-based data from the MPD to train the SVM model for predicting the total energy per atom, formation energy per atom and PF of the Si-based alloy structures. The focus lies in selecting suitable structural descriptors that enable the model to achieve the best transferability, high prediction accuracies, and least dependence on data quantity. Three common descriptors are employed to convert atomic structures into ML inputs, and the model's performance is evaluated on completely new Si alloys. We demonstrate that the SVR model exhibits high predictive accuracy for Si alloys compared to more advanced approaches that require extensive data generation through ab-initio simulations for training[67,68].

## 2. Methodology

### 2.1 Dataset Preparation:

The dataset used in this study consists of 745 inorganic structures comprising Si and $A_xSi_y$ alloys, where A represents elements such as Li, Na, K, Mg, Ca, and Al. The stoichiometric ratios of A and Si are denoted by $x$ and $y$, respectively. The dataset includes associated properties such as total energy per atom, packing fraction, and formation energy per atom. In MPD, researchers widely utilize DFT calculations based on the ground state energy to determine the total energy of compounds[69]. They evaluate the thermodynamic stability of a compound by considering Gibbs Free Energy ($\Delta G$) while utilizing change in enthalpy ($\Delta H$) as a practical approximation[70]. This simplification effectively equates ΔG and ΔH to the change in internal energy (ΔU), enabling the use of total internal energy as an approximation for thermodynamic stability at absolute zero temperature (0 K)[70]. A compound's formation energy ($\Delta H_f$) is determined by considering the ground state energies of its constituent elements and the combination itself [71]. It quantifies the difference between the total enthalpy of the compound and the sum of the enthalpies of the constituent elements, considering their stoichiometric fractions[70]. Notably, cohesive energy and formation energy share a relationship, where the cohesive energy represents energy release and the formation energy represents energy consumption, with opposite signs to reflect bond formation or breakup[71]. To retrieve the dataset, we utilized the MPD and accessed it through the application programming interface (API) using Matminer[72] and Python Materials Genomics (pymatgen)[73] libraries in python. Python dataset extraction implementation available on GitHub (https://github.com/joy1303125/Si-based-Material-stability-prediction/tree/main).

Figure 1 shows a sample structure from each $A_xSi_y$ alloy, representing the Si alloys with different elements A. In the dataset, all the metals (A) are either monovalent or bivalent, except for the case of Al. We used this dataset to validate the transferability of the ML model to multivalent $A_xSi_y$ alloys. Specifically, 12 $Na_xSi_y$ and 3 $Al_xSi_y$ structures are considered as the test dataset, while the remaining dataset as our training dataset. Additional information regarding the test structures can be found in Table S7.

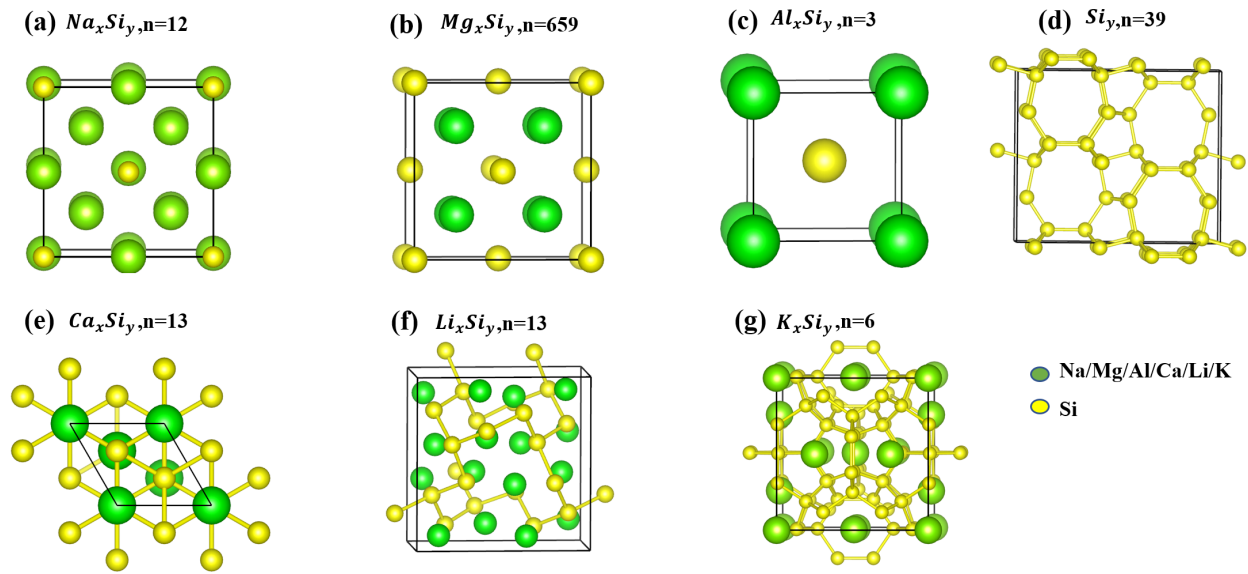

**Figure 1** Representative atomic structures from the dataset consisting of inorganic crystal structures of Si and $A_xSi_y$ alloys, where A= Li, Na, K, Mg, Ca and Al, x and y represents stoichiometric ratio of A and Si. List of Si based metal anode structures are represented as **(a)** $Na_3Si$, **(b)** $Mg_2Si$, **(c)**AlSi, **(d)** Si, **(e)**$CaSi_2$, **(f)** LiSi, **(g)** $K_4Si_{23}$ where *n* is the total number of sample structures present for each alloy category.

## 2.2 Descriptors

The key to any ML model's success is the rightful representation of the atomic structures. The choice of the descriptors is sensitive to the learning labels and the model's paradigm. Traditionally, the process of selecting suitable descriptors involves a trial-and-error[46] approach. In this study, we employ multiple descriptors to represent atomic structures and evaluate their performance in predicting the stability metrices of the structures. The three descriptors compared in the study are X-ray diffraction pattern (XRD), sine coulomb matrix (SCM), and orbital field matrix (OFM). The primary characteristics of each of these descriptors have been detailed in the supplementary section.

## 2.3 Support Vector Regression

SVR is a regression model that utilizes a statistical learning approach to forecast continuous values. It fits a hyperplane to the data in n-dimensional space and employs Vapnik's insensitive region approach to create a generalized model with high prediction ability[74]. SVR has several hyperparameters to choose, such as the tube width (epsilon), suitable kernels (linear, polynomial, or radial basis function), the regularization parameter C, and gamma. The appropriate selection of these parameters requires applying additional hyperparameter tuning techniques. Two different types of hyperparameter tuning techniques are Grid Search CV and Bayes Search CV. Both techniques can be implemented in Python using the scikit-learn library[75]. The model can be evaluated using the repeated K-fold cross-validation technique to avoid overfitting. In this work, SVR algorithm has been implemented on python library called scikit-learn[75]. The training data is divided into a 5-fold cross-validation dataset, where 1-fold is considered as a validation dataset, and the rest of the fold is going onto training the model. Each fold result is repeated 10 times to keep away from the noise in the predictions. Two different optimization techniques, Grid Search CV and Bayes Search CV, have been used for the best hyperparameters of our SVR model. Details of these two hyperparameters are detailed in the supplementary section (see section 2.1 and 2.2).

## 3. Results and Discussions

In this section, we discuss our model's exploratory data analysis and performance for the test dataset of Si alloys described in section 2.1. Prediction results of system total energy and packing fraction (PF) for validation and test data are compared using root mean square error (RMSE) value. In addition, model's prediction ability is tested for 3 different types of structural descriptor detailed in the supplementary section 1(see Supporting Information section 1.1-1.3).

### 3.1 Exploratory Data Analysis

We analyzed the data distribution before fitting the model to interpret trends in mean, variance, frequency, and outliers. For interpreting the complete data, violin plots are used that come within seaborn python library [76]. The violin plot displays the inner interquartile range as a thick black box and the median value as a white dot. In Figure 2, violin plots depict the data pattern of output labels in the $A_xSi_y$ dataset, namely total energy/atom, PF, and formation energy/atom. From Figure 2a,b, it is evident that $K_xSi_y$ and $Na_xSi_y$ structures exhibit the highest variance for both output labels, total energy/atom and PF, compared to the remaining data ($Li_xSi_y$, $Mg_xSi_y$, $Ca_xSi_y$, $Al_xSi_y$ and $Si_y$). The protruding plots beyond the interquartile range for $Na_xSi_y$ and $Mg_xSi_y$ structures in figure 2a, b suggests the presence of outliers in the data. In figure 2c, for the formation energy per atom data, we observe outliers in $Mg_xSi_y$ structures compared to the rest of the dataset. We test all our results by removing outliers followed by the modified Z score method[77] (shown in Table S4-S6). Regarding performance, the cross-validation dataset exhibits improved outcomes compared to the original data points. However, when evaluating the test dataset, we observe a degradation in performance (as depicted in Tables 1-3, S4-S6).

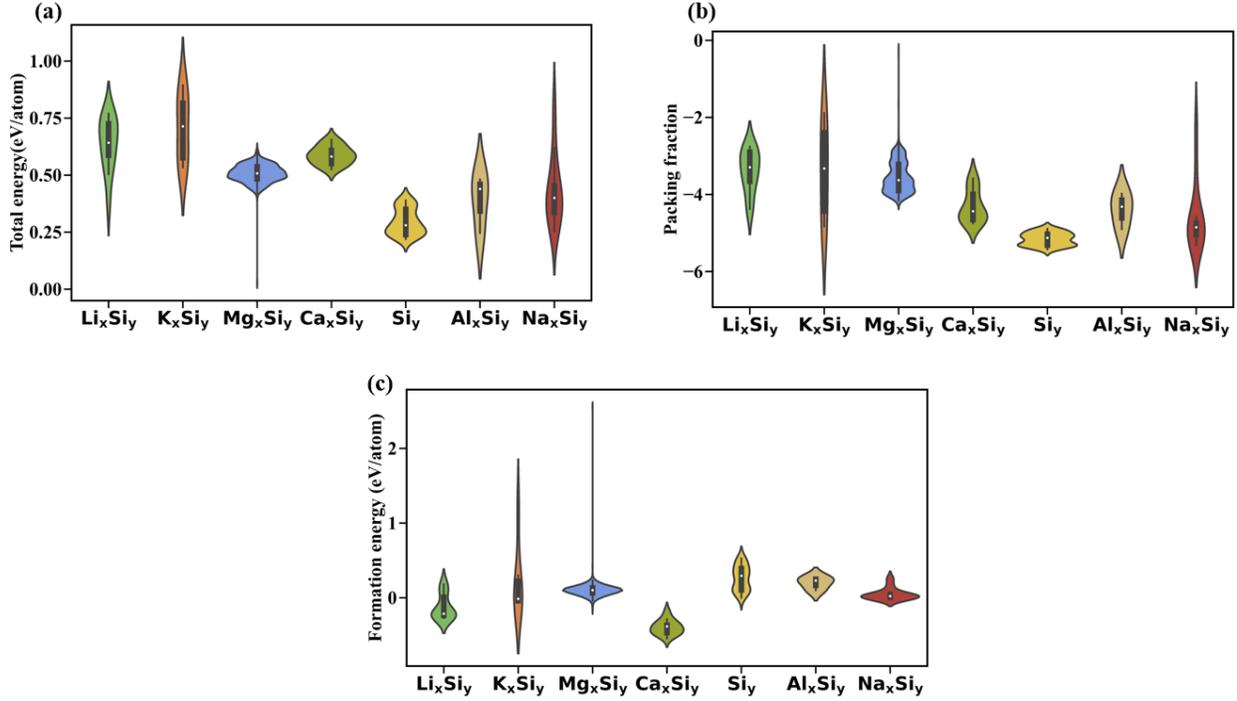

**Figure 2** Data distribution of Si and $A_xSi_y$ alloy structures based on **(a)** Total energy/atom(eV/atom) **(b)** Packing fraction **(c)** Formation energy/atom. Black bar represents the interquartile range, white dot shows the median value and data that falls outside the interquartile range is considered as outliers.

### 3.2 Model Performance

To access the prediction capability of the SVR model, the model is fitted on training data as described in section 2.1. Each training data is further sectioned into training and validation datasets based on the 5-fold cross validation method, which separates 20% of the total training datapoints for the model validation during the training. The RMSE between the predicted and actual output values is employed to measure the model's accuracy. The model performance of the two trained models is evaluated by predicting the total energy/atom, formation energy/atom and PF values for $Na_xSi_y$ and $Al_xSi_y$ structures, respectively. The performance of trained SVR models is detailed in sections 3.2.1 and 3.2.2 below.

### 3.2.1 Test dataset study on $Na_xSi_y$ and $Al_xSi_y$ structures for total energy and PF prediction:

In the first experiment, SVR is trained on 730 structures of $A_xSi_y$ alloys, where A = Li, K, Mg, and Ca. Hyperparameter tuning for SVR is performed using both Grid Search CV and Bayes Search CV, resulting in different sets of hyperparameter values, which are tabulated in Table 1 and Table 2. For each of these hyperparameter sets, three SVR models are trained with different structural input descriptors, as detailed in section 2.2. In total, six SVR models are trained, incorporating various combinations of hyperparameters and structural descriptor methods. The trained models are used to predict the total energy/atom and PF of 12 $Na_xSi_y$ and 3 $Al_xSi_y$ structures as test datapoints. Figure 3 describes the results obtained during the validation and testing of 6 models. The obtained RMSE value for total energy/atom prediction and PF prediction are plotted for validation (see Figure 3(a, b)) and test dataset (see Figure 3(c, d)).

From the histograms in Figure 3(a, b), it is visible that RMSE values for validation using Grid search and Bayes Search are nearly identical, except for the total energy/atom prediction using XRD descriptor in Figure 3a, which shows validation results for total energy per atom prediction. Among the different descriptor methods, the highest RMSE of 0.73 eV/atom was obtained when using XRD descriptors with Bayes Search CV parameters. On the other hand, the lowest RMSE of 0.15 eV/atom was achieved when employing OFM descriptors with Bayes Search CV parameters. Regarding the PF prediction in validation dataset, the lowest RMSE of 0.03 was obtained when using both XRD and OFM descriptors. Conversely, the highest RMSE of 0.03 was observed with Sine descriptors (Figure 3b).

In test datapoints, the XRD-Grid search-based model showed the lowest RMSE of 0.28 eV/atom for total energy per atom prediction (Figure 3c). The highest RMSE of 1.16 eV/atom was observed with Sine descriptors. Regarding PF prediction in test datapoints, the XRD-bayes model achieved the lowest RMSE of 0.07, while the highest RMSE of 0.20 was obtained with Sine descriptors (Figure 3d). These results emphasize the importance of selecting appropriate descriptors and optimization strategies for accurate predictions of total energy per atom and PF.

A comprehensive comparison of the performance of hyperparameter tuning methods and structural descriptors for total energy/atom and PF are presented in Table 1 and Table 2, respectively. Additionally, Figure 1b illustrates that the datasets exhibit a high degree of skewness on $Mg_xSi_y$ dataset. Consequently, 371 data points have been excluded by considering energy above hull greater than 20 eV/atom, as they fall outside the metastable material range[78]. However, the performance in terms of the test and validation datasets has decreased for all cases, as indicated in Tables S1, and S2. Therefore, we have decided to utilize the unfiltered dataset for our test and validation cases when predicting total energy per atom and PF.

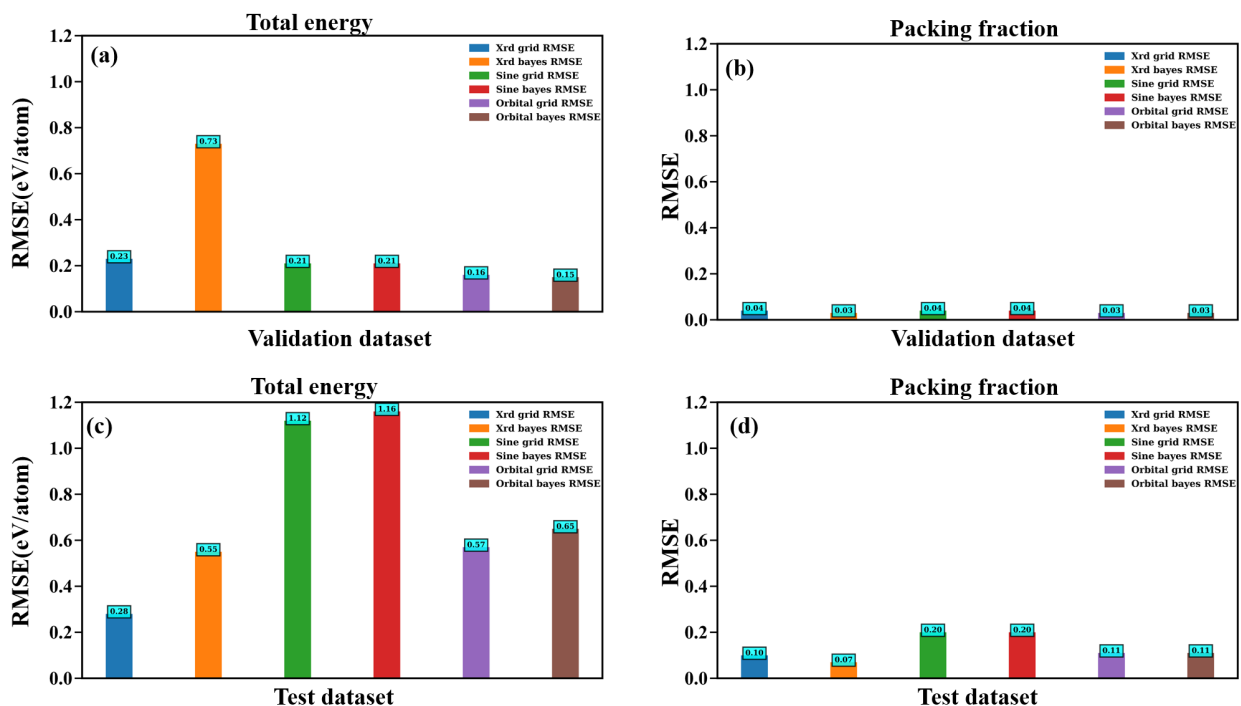

**Figure 3** Performance comparison of hyperparameter tuning methods Grid Search CV and Bayes Search CV. **(a)** RMSE value of total energy/atom for validation dataset, **(b)** RMSE value of packing fraction for validation dataset, **(c)** RMSE value of total energy/atom for the test dataset ($Na_xSi_y$ and $Al_xSi_y$), **(d)** RMSE value of packing fraction for test dataset ($Na_xSi_y$ and $Al_xSi_y$). Three different structural descriptors with two different optimizations techniques generated 6 set of results labeled on top right side.

Figure 4 illustrates the predicted total energy/atom and PF for test $Na_xSi_y$ and $Al_xSi_y$ structures, utilizing the SVR models based on Grid Search CV and Bayes Search CV. For comparison, the actual values of total energy/atom and packing fraction are depicted as red scatter plots in Figure 4(a-f). The total energy predictive supremacy of the XRD descriptor is evident in Figure 4a. Regarding the packing fraction prediction, Figure 4d demonstrates that the XRD-Bayes model outperforms the other descriptors (Figure 4e, f). The RMSE for PF of test $Na_xSi_y$ and $Al_xSi_y$ structures ranges from 0.07 to 0.11 for all XRD and OFM models (Figure 3d), as noted in Table 2. Hence, the findings from Figure 4d-f and Table 1 provide compelling evidence of the superior performance of XRD and OFM descriptors over SCM descriptors in predicting both total energy per atom and PF.

Although the implementation of Bayes Search CV and Grid Search CV with SVR yields similar energy and structural predictions, we emphasize that Bayes Search CV is faster than the Grid Search CV. Grid Search CV iterates over complete permutated combinations of hyperparameters, which takes 144 hours on 32 cores. In contrast, Bayes search CV completes the search in just 0.08 hours on the same computational facility. Therefore, we can conclude that Bayes Search CV is nearly 1800 times more efficient than Grid Search CV based on the total execution time of the two approaches.

**Comparison of Grid search CV and bayes search CV for XRD descriptors**

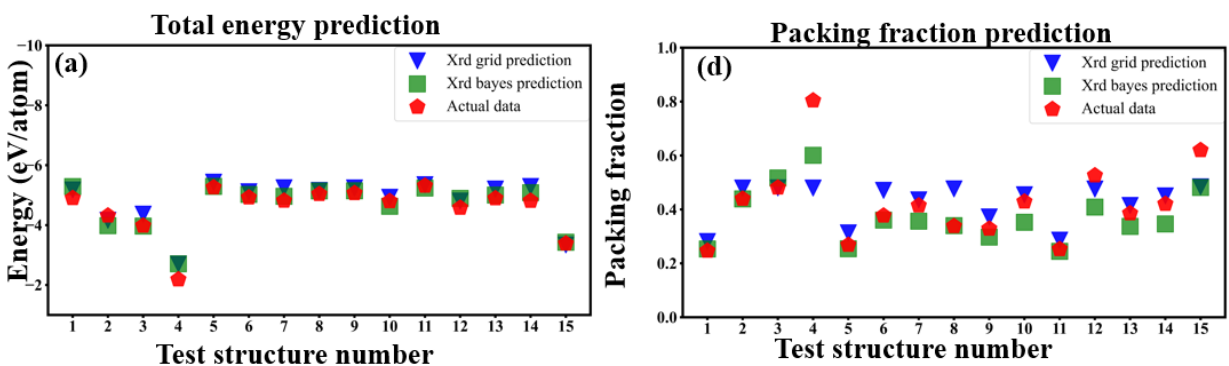

**Comparison of Grid search CV and bayes search CV for Sine descriptors**

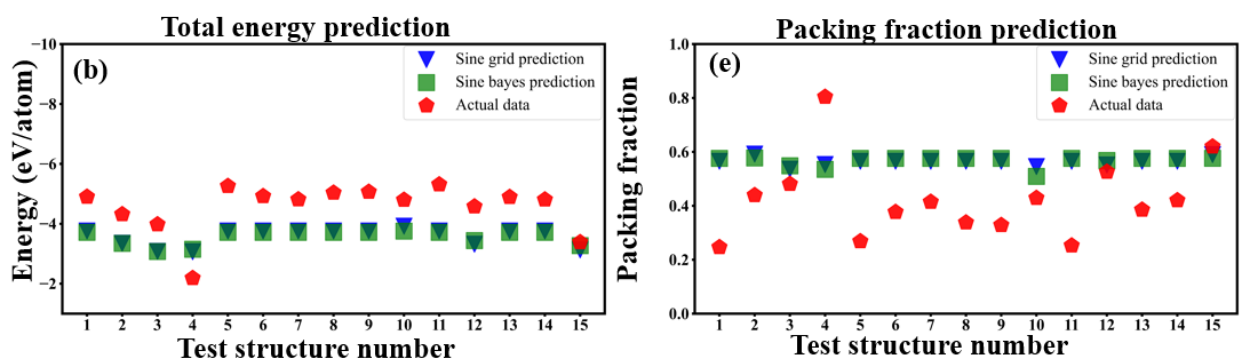

**Comparison of Grid search CV and bayes search CV for OFM descriptors**

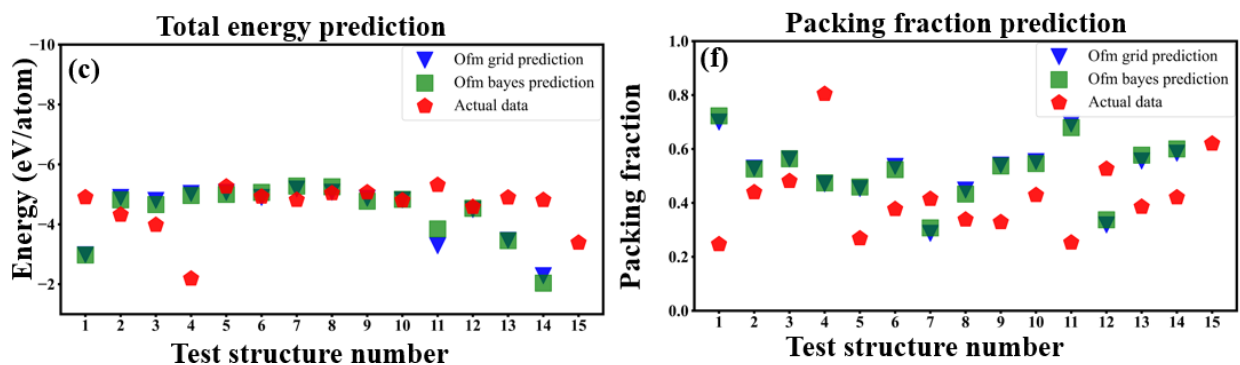

**Figure 4** Comparing predictions using three structural descriptors with different hyperparameter search approaches for $Na_xSi_y$ and $Al_xSi_y$ test dataset. **(a, b, c)** Total energy/atom of the test structures, **(d, e, f)** Packing fraction of the test structures.

**Table 1** SVR hyperparameters, descriptor used and associated results for $Na_xSi_y$ and $Al_xSi_y$ test dataset for total energy/atom prediction.

| Descriptor Name | Hyperparameter Tuning Technique Name | C | Gamma | Epsilon | Train RMSE(eV/atom) | Test RMSE(eV/atom) | Validation RMSE(eV/atom) |
|---|---|---|---|---|---|---|---|
|  | Grid Search CV | 51 | 0.0001 | 5.00E-05 | 0.17 | 0.28 | 0.23 |
| XRD | Bayes Search CV | 600 | 1.38E-06 | 1.56E-06 | 0.55 | 0.55 | 0.73 |
|  | Grid Search CV | 21 | 0.05 | 0.1 | 0.1 | 1.12 | 0.21 |
| Sine | Bayes Search CV | 4.98 | 0.15 | 0.02 | 0.1 | 1.16 | 0.21 |
|  | Grid Search CV | 11 | 0.01 | 0.0005 | 0.13 | 0.57 | 0.16 |
| OFM | Bayes Search CV | 276.76 | 0.009 | 0.008 | 0.08 | 0.65 | 0.15 |

**Table 2** SVR hyperparameters, descriptor used and associated results for $Na_xSi_y$ and $Al_xSi_y$ test dataset for PF prediction.

| Descriptor Name | Hyperparameter Tuning Technique Name | C | Gamma | Epsilon | Train RMSE | Test RMSE | Validation RMSE |
|---|---|---|---|---|---|---|---|
|  | Grid Search CV | 1 | 0.1 | 0.005 | 0.004 | 0.1 | 0.04 |
| XRD | Bayes Search CV | 600 | 2.06E-05 | 3.00E-06 | 0.01 | 0.07 | 0.03 |
|  | Grid Search CV | 1 | 0.1 | 0.005 | 0.02 | 0.2 | 0.04 |
| Sine | Bayes Search CV | 0.96 | 0.08 | 0.02 | 0.02 | 0.2 | 0.04 |
|  | Grid Search CV | 1 | 0.01 | 5.00E-05 | 0.022 | 0.11 | 0.03 |
| OFM | Bayes Search CV | 0.43 | 7.00E-03 | 2.00E-04 | 0.02 | 0.11 | 0.03 |

### 3.2.2 Test Dataset study on $Na_xSi_y$ and $Al_xSi_y$ structures for formation energy/atom prediction:

The histograms in Figure 5a, b depict the RMSE values for both the validation and test datasets using Grid search and Bayes search, with the exception of the grid search CV parameter-based OFM test dataset case (see Figure 5b). Figure 5a demonstrates that the best RMSE of 0.12 eV/atom is achieved when utilizing both XRD and OFM descriptors for the validation dataset. On the other hand, Figure 5b reveals that the XRD-Bayes model achieves the lowest RMSE value of 0.08 eV/atom, while Sine descriptor perform poorly, yielding an RMSE value of 0.17 eV/atom. Table 3 provides a comprehensive comparison of the performance of hyperparameter tuning methods and structural descriptors for formation energy/atom.

Figure 6(a, b, c) further supports the evaluation of which descriptor correctly predicts the actual value of formation energy/atom. The errors between predicted values and actual values are lowest for formation energy/atom when structures are described by XRD-bayes, as shown in Figure 6a. Conversely, Figure 6b illustrates the poor performance of the SCM descriptor in predicting actual formation energy/atom.

Due to heavy skewness towards $Mg_xSi_y$ structure in the dataset (Figure 1b), we excluded 371 data points with energy above hull > 20 eV/atom[78]. All predictions for formation energy/atom were carried out using Bayes search. The performance on the validation dataset exhibited a consistent RMSE. However, for the test dataset, the model's prediction ability deteriorates, resulting in an increased RMSE value of 0.11 eV/atom (see Table S3), where the best RMSE value obtained from Figure 5b is 0.08 eV/atom. This represents a 37.5% increase in test dataset error compared to the RMSE value obtained with the actual data points. Therefore, original dataset has the advantage of using it as a training data point as they help cope with the limited availability of data. The hyperparameters and performance in terms of training, testing, and validation datasets for both the original and filtered datasets are provided in Table 3 and Table S3.

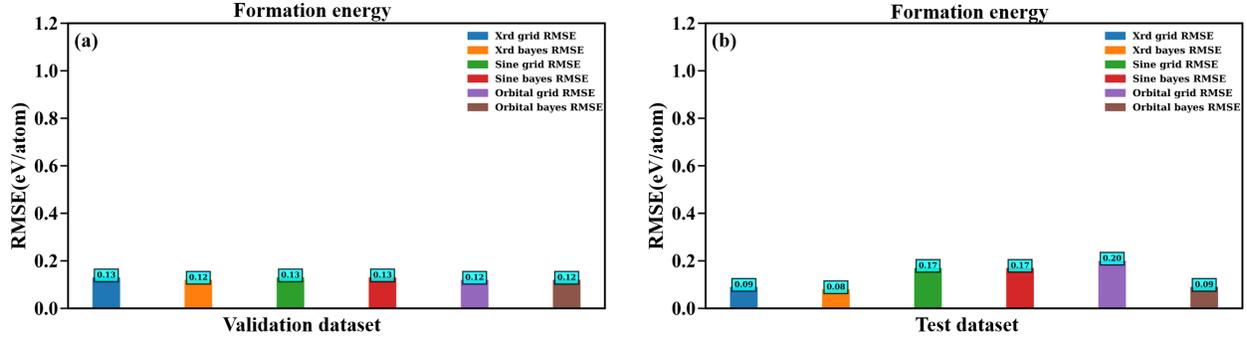

**Figure 5** Performance comparison of hyperparameter tuning methods Grid Search CV and Bayes Search CV. RMSE value of formation energy/atom for **(a)** validation dataset, and **(b)** test dataset ($Na_xSi_y$ and $Al_xSi_y$). Three different structural descriptors with two different optimization technique generated 6 set of results labeled on top right side.

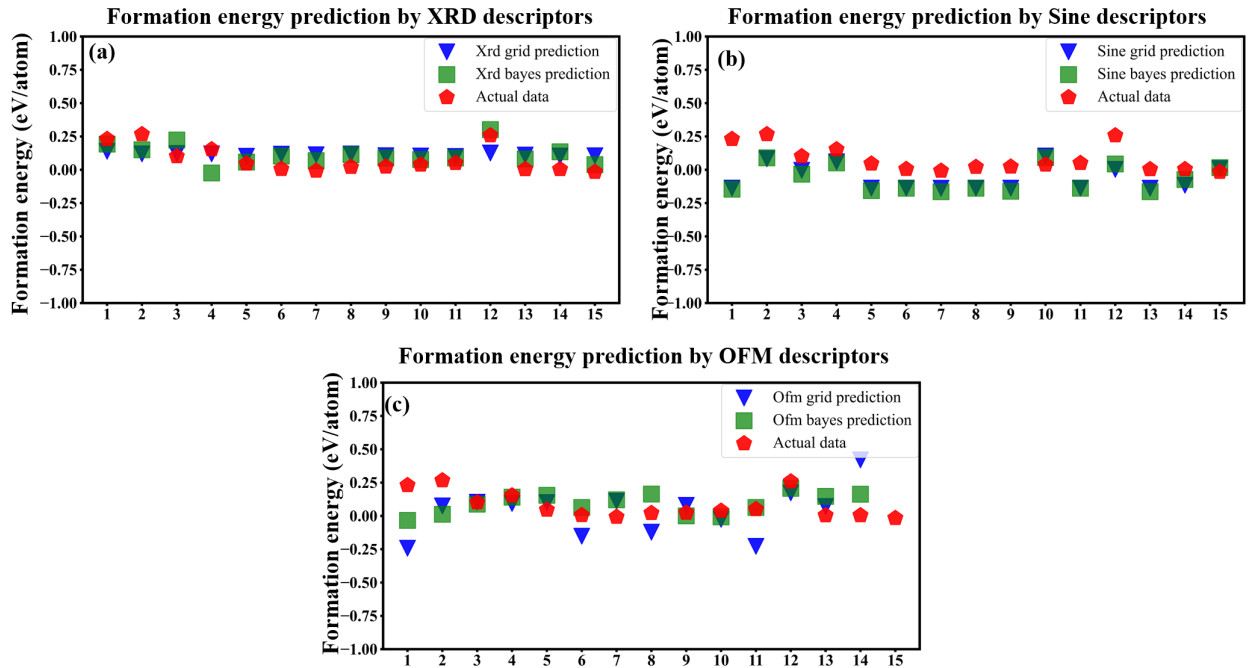

**Figure 6** Comparing predictions using three structural descriptors with different hyperparameter search approaches for $Na_xSi_y$ and $Al_xSi_y$ test dataset. (a, b, c) Formation energy/atom of the test structures. XRD descriptors prediction value matches with the actual data points (see Figure 6a).

All the test structures and their properties have been presented on the supplementary information (see Table S7).

**Table 3** SVR hyperparameters, descriptor used and associated results for $Na_xSi_y$ and $Al_xSi_y$ test dataset for formation energy/atom prediction.

| Descriptor Name | Hyperparameter Tuning Technique Name | C | Gamma | Epsilon | Train RMSE(eV/atom) | Test RMSE(eV/atom) | Validation RMSE(eV/atom) |
|---|---|---|---|---|---|---|---|
| | Grid Search CV | 1 | 0.01 | 0.05 | 0.06 | 0.09 | 0.13 |
| XRD | Bayes Search CV | 103.21 | 9.00E-04 | 0.07 | 0.04 | 0.08 | 0.12 |
| | Grid Search CV | 1 | 0.01 | 0.05 | 0.12 | 0.17 | 0.13 |
| Sine | Bayes Search CV | 1.73 | 5.00E-03 | 1.72E-05 | 0.12 | 0.17 | 0.13 |
| | Grid Search CV | 201 | 0.01 | 0.005 | 0.07 | 0.2 | 0.12 |
| OFM | Bayes Search CV | 5.64 | 0.5 | 0.007 | 0.06 | 0.09 | 0.12 |

Similar work has been done on predicting formation energy/atom, where the Kernel Ridge Regression (KRR) model is trained on 11674 material structures collected from MPD [72]. The study reported the RMSE of formation energy/atom prediction for validation datapoints to be 0.10 eV/atom. This performance is comparable to our current SVM-Bayes model, which achieved an RMSE of 0.12 eV/atom on the validation dataset when trained with only 730 structural data (see Figure 7). Additionally, another similar study demonstrated that increasing the dataset size has a positive impact on prediction accuracy[79]. They employed a CNN-OFM based model and observed that as the dataset size increased from 750 to 4,000, the RMSE for formation energy per atom on the validation dataset decreased from 0.18 eV/atom to 0.10 eV/atom (see Figure 7). Figure 7 compares the errors noted in previously reported energy prediction ML models and the training data used against the presented SVR-Bayes-XRD model. However, by excluding 21 outliers from the training data points following the modified Z score method[77], we can obtain an RMSE value of 0.04 eV/atom, shown in the last bar of Figure 7 and Table S4 . This comparison demonstrates that with the qualitative training data selection, hyperparameter tuning methods, and use of appropriate structural descriptors, ML models can overcome the need for extensive data requirements for training and accurate predictions.

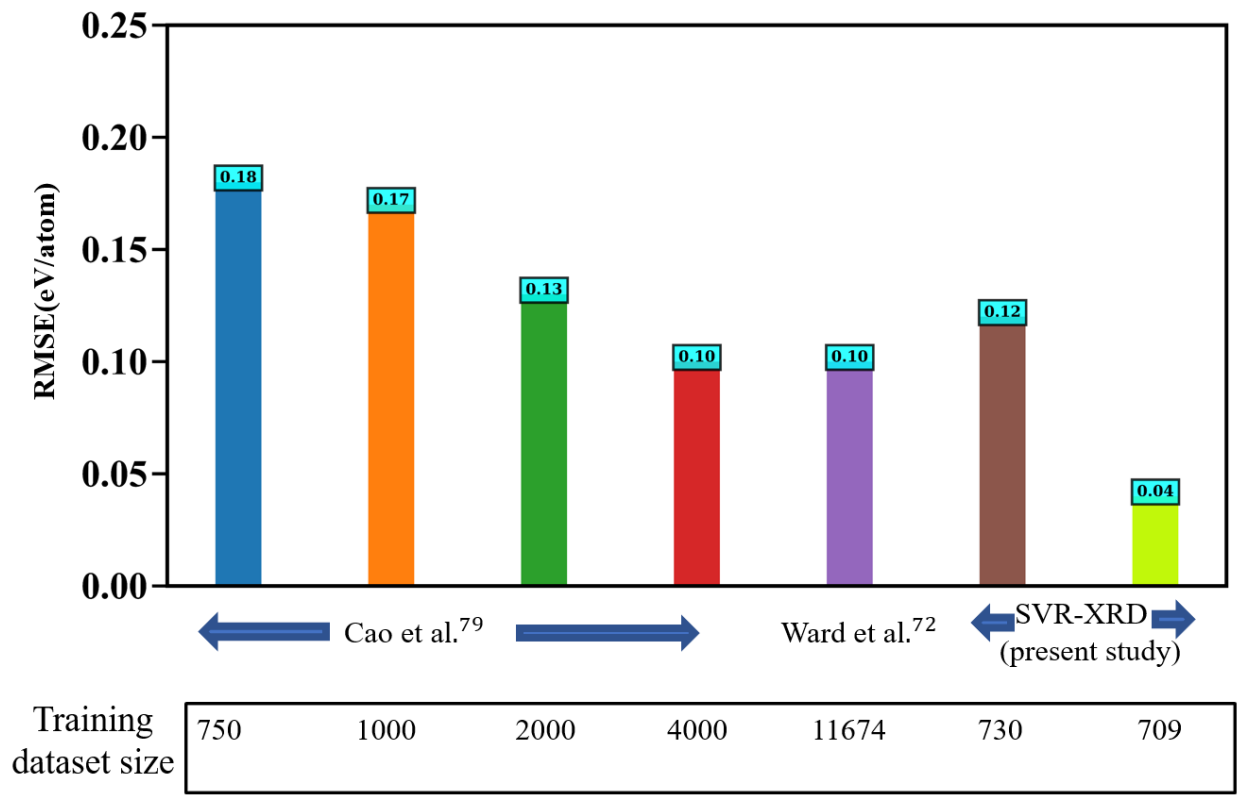

**Figure 7** Performance comparison of the presented SVR-Bayes-XRD model with previously reported[72,79] ML based energy prediction models in terms of RMSE value and training dataset size. SVR-Bayes-XRD model shows better performance with limited training data size in comparison to previous reports.

## 4. Conclusion

In summary, we propose SVR based machine learning method to speedily predict the thermodynamic and structural stability of Si alloying anodes before experimental design. The use of hyperparameter tuning methods such as Grid Search CV and Bayes Search CV, and the structural descriptors to convert atomic coordinates to comprehensive machine learning inputs have been elaborated. The predictive ability of three different types of descriptors has been studied for $A_xSi_y$ atomic systems. XRD descriptor of the $A_xSi_y$ structures as input data for the SVR model performed most reliably, especially considering the training structures were a mix of crystal, amorphous and different electronic configuration systems. While the OFM descriptor predicted total energy/atom, formation energy/atom and packing fraction with the lowest errors and highest accuracies for similar electronic configurations, OFM failed for the test cases where electronic configurations were slightly different from the training data (valency of cations). These results demonstrate that the choice of descriptor has more weight than the training data in making an ML model transferable to new systems. Moreover, the prediction accuracies were improved by the coupled use of SVR with Grid search CV method. In the two demonstrated experiments, hyperparameter selection by the Grid search CV method showed better predictions for the new structures. Though training and prediction times were shorter for SVR coupled with the Bayes search CV method, SVR-Bayes approach is suitable for predicting the stability of similar structures where transferability is not targeted. This study attempts to establish that the requirements of large datasets for machine learning based approaches in material science domain can be overcome with the qualitative selection of training data, hyperparameter tuning methods, and appropriate structural descriptors.

# AUTHOR INFORMATION

## Corresponding Authors

Vidushi Sharma, Ph.D., was formerly at New Jersey Institute of Technology during the project, and is now at IBM Almaden Research Center, San Jose, CA 95120, USA.

Email: vs574@njit.edu, vidushis@ibm.com

## Author Contributions

V.S. and J.K. conceived the project. J.K. performed all work and wrote the manuscript with V.S and D.D.  All authors approved the final version of the manuscript.

## CONFLICT OF INTEREST OR COMPETING INTEREST

The authors have no conflicts of interest to declare. All authors have seen and agree with the contents of the manuscript and there is no financial interest to report. We certify that the submission is original work and is not under review at any other publication.

## ACKNOWLEDGEMENT

The work is supported by National Science Foundation (NSF), Award Number # 2126180. Authors acknowledge Advanced Cyberinfrastructure Coordination Ecosystem: Service & Support (ACCESS) for the computational facilities (Award Number – DMR180013).

## DATA AND CODE AVAILABILITY

The data, pre- and post-processing code reported in this paper is available on GitHub (https://github.com/joy1303125/Si-based-Material-stability-prediction)

**SUPPLEMENTARY INFORMATION**

Descriptors used in the study; Hyperparameter optimization methods; Prediction results with training data excluding high energy $Mg_xSi_y$ structures; Prediction results with training data excluding outliers; Test Structure Information

**ETHICAL APPROVAL**

Not Applicable

**REFERENCES**

(1)     Nishi, Y. Lithium Ion Secondary Batteries; Past 10 Years and the Future. *J. Power Sources* **2001**, *100* (1–2), 101–106. https://doi.org/10.1016/S0378-7753(01)00887-4.

(2)     Shenoy, V. B.; Johari, P.; Qi, Y. Elastic Softening of Amorphous and Crystalline Li – Si Phases with Increasing Li Concentration : A First-Principles Study. *J. Power Sources* **2010**, *195* (19), 6825–6830. https://doi.org/10.1016/j.jpowsour.2010.04.044.

(3)     Limthongkul, P.; Jang, Y. Il; Dudney, N. J.; Chiang, Y. M. Electrochemically-Driven Solid-State Amorphization in Lithium-Silicon Alloys and Implications for Lithium Storage. *Acta Mater.* **2003**, *51* (4), 1103–1113. https://doi.org/10.1016/S1359-6454(02)00514-1.

(4)     Grosjean, C.; Herrera Miranda, P.; Perrin, M.; Poggi, P. Assessment of World Lithium Resources and Consequences of Their Geographic Distribution on the Expected Development of the Electric Vehicle Industry. *Renew. Sustain. Energy Rev.* **2012**, *16* (3), 1735–1744. https://doi.org/10.1016/j.rser.2011.11.023.

(5)     Kubota, K.; Dahbi, M.; Hosaka, T.; Kumakura, S.; Komaba, S. Towards K-Ion and Na-

Ion Batteries as "Beyond Li-Ion." *Chem. Rec.* **2018**, *18* (4), 459–479. https://doi.org/10.1002/tcr.201700057.

(6)     Arrieta, U.; Katcho, N. A.; Arcelus, O.; Carrasco, J. First-Principles Study of Sodium Intercalation in Crystalline Na x Si24 (0 ≤ x ≤ 4) as Anode Material for Na-Ion Batteries. *Sci. Rep.* **2017**, *7* (1), 1–8. https://doi.org/10.1038/s41598-017-05629-x.

(7)     Legrain, F.; Malyi, O. I.; Manzhos, S. Comparative Computational Study of the Energetics of Li, Na, and Mg Storage in Amorphous and Crystalline Silicon. *Comput. Mater. Sci.* **2014**, *94* (C), 214–217. https://doi.org/10.1016/j.commatsci.2014.04.010.

(8)     Ponrouch, A.; Tchitchekova, D.; Frontera, C.; Bardé, F.; Dompablo, M. E. A. De; Palacín, M. R. Assessing Si-Based Anodes for Ca-Ion Batteries: Electrochemical Decalciation of CaSi2. *Electrochem. commun.* **2016**, *66*, 75–78. https://doi.org/10.1016/j.elecom.2016.03.004.

(9)     Zhang, D.; Fu, J.; Wang, Z.; Wang, L.; Corsi, J. S.; Detsi, E. Perspective—Reversible Magnesium Storage in Silicon: An Ongoing Challenge. *J. Electrochem. Soc.* **2020**, *167* (5), 050514. https://doi.org/10.1149/1945-7111/ab736b.

(10)    Beaulieu, L. Y.; Hatchard, T. D.; Bonakdarpour, A.; Fleischauer, M. D.; Dahn, J. R.; Soc, J. E.; A-a, P.; Beaulieu, L. Y.; Hatchard, T. D.; Bonakdarpour, A.; Fleischauer, M. D. Reaction of Li with Alloy Thin Films Studied by In Situ AFM Service Reaction of Li with Alloy Thin Films Studied by In Situ AFM. **2003**, *150* (11). https://doi.org/10.1149/1.1613668.

(11)    Lee, S.; Lee, J.; Chung, S.; Lee, H.; Lee, S.; Baik, H. Stress Effect on Cycle Properties of the Silicon Thin- ® Lm Anode. **2001**, *c*, 191–193.

(12)    Wang, W.; Kumta, P. N. Reversible High Capacity Nanocomposite Anodes of Si / C /

SWNTs for Rechargeable Li-Ion Batteries. **2007**, *172*, 650–658. https://doi.org/10.1016/j.jpowsour.2007.05.025.

(13)   Kim, H.; Chou, C.; Ekerdt, J. G.; Hwang, G. S. Structure and Properties of Li - Si Alloys : A First-Principles Study. **2011**, 2514–2521.

(14)   Wan, W.; Zhang, Q.; Cui, Y.; Wang, E. First Principles Study of Lithium Insertion in Bulk Silicon. *J. Phys. Condens. Matter* **2010**, *22* (41). https://doi.org/10.1088/0953-8984/22/41/415501.

(15)   Chevrier, V. L.; Zwanziger, J. W.; Dahn, J. R. First Principles Study of Li-Si Crystalline Phases: Charge Transfer, Electronic Structure, and Lattice Vibrations. *J. Alloys Compd.* **2010**, *496* (1–2), 25–36. https://doi.org/10.1016/j.jallcom.2010.01.142.

(16)   Chiang, H. H.; Lu, J. M.; Kuo, C. L. First-Principles Study of the Structural and Dynamic Properties of the Liquid and Amorphous Li-Si Alloys. *J. Chem. Phys.* **2016**, *144* (3). https://doi.org/10.1063/1.4939716.

(17)   Fan, F.; Huang, S.; Yang, H.; Raju, M. Mechanical Properties of Amorphous Li x Si Alloys : A Reactive Force Field Study. *074002*. https://doi.org/10.1088/0965-0393/21/7/074002.

(18)   Johari, P.; Qi, Y.; Shenoy, V. B. The Mixing Mechanism during Lithiation of Si Negative Electrode. **2011**, 5494–5500.

(19)   Lee, S.; Ko, M.; Jung, S. C.; Han, Y. K. Silicon as the Anode Material for Multivalent-Ion Batteries: A First-Principles Dynamics Study. *ACS Appl. Mater. Interfaces* **2020**, *12* (50), 55746–55755. https://doi.org/10.1021/acsami.0c13312.

(20)   Ponrouch, A.; Tchitchekova, D.; Frontera, C.; Bardé, F.; Arroyo-de Dompablo, M. E.; Palacín, M. R. Assessing Si-Based Anodes for Ca-Ion Batteries: Electrochemical

Decalciation of CaSi2. *Electrochem. commun.* **2016**, *66*, 75–78.

(21)    Niu, J.; Zhang, Z.; Aurbach, D. Alloy Anode Materials for Rechargeable Mg Ion Batteries. *Adv. Energy Mater.* **2020**, *10* (23), 2000697.

(22)    Mandal, S.; Haule, K.; Rabe, K. M.; Vanderbilt, D. Systematic Beyond-DFT Study of Binary Transition Metal Oxides. *npj Comput. Mater.* **2019**, *5* (1), 1–8.

(23)    Li, W.; Walther, C. F. J.; Kuc, A.; Heine, T. Density Functional Theory and beyond for Band-Gap Screening: Performance for Transition-Metal Oxides and Dichalcogenides. *J. Chem. Theory Comput.* **2013**, *9* (7), 2950–2958.

(24)    Olson, J.; Priester, M.; Luo, J.; Chopra, S.; Zieve, R. J. Packing Fractions and Maximum Angles of Stability of Granular Materials. *Phys. Rev. E - Stat. Nonlinear, Soft Matter Phys.* **2005**, *72* (3), 1–6. https://doi.org/10.1103/PhysRevE.72.031302.

(25)    He, Q.; Yu, B.; Li, Z.; Zhao, Y. Density Functional Theory for Battery Materials. *Energy Environ. Mater.* **2019**, *2* (4), 264–279.

(26)    Deringer, V. L. Modelling and Understanding Battery Materials with Machine-Learning-Driven Atomistic Simulations Journal of Physics : Energy OPEN ACCESS Modelling and Understanding Battery Materials with Machine-Learning-Driven Atomistic Simulations. *J. Phys. Energy* **2020**, *2*, 041003-1–11.

(27)    Deringer, V. L.; Bernstein, N.; Bartók, A. P.; Cliffe, M. J.; Kerber, R. N.; Marbella, L. E.; Grey, C. P.; Elliott, S. R.; Csányi, G. Realistic Atomistic Structure of Amorphous Silicon from Machine-Learning-Driven Molecular Dynamics. *J. Phys. Chem. Lett.* **2018**, *9* (11), 2879–2885. https://doi.org/10.1021/acs.jpclett.8b00902.

(28)    He, X.; Zhu, Y.; Epstein, A.; Mo, Y. Statistical Variances of Diffusional Properties from Ab Initio Molecular Dynamics Simulations. *Npj Comput. Mater.* **2018**, *4* (1), 1–9.

(29)    Xie, T.; Grossman, J. C. Crystal Graph Convolutional Neural Networks for an Accurate and Interpretable Prediction of Material Properties. *Phys. Rev. Lett.* **2018**, *120* (14), 145301. https://doi.org/10.1103/PhysRevLett.120.145301.

(30)    Sanyal, S.; Balachandran, J.; Yadati, N.; Kumar, A.; Rajagopalan, P.; Sanyal, S.; Talukdar, P. MT-CGCNN: Integrating Crystal Graph Convolutional Neural Network with Multitask Learning for Material Property Prediction. **2018**.

(31)    Karamad, M.; Magar, R.; Shi, Y.; Siahrostami, S.; Gates, I. D.; Farimani, A. B. Orbital Graph Convolutional Neural Network for Material Property Prediction. *Phys. Rev. Mater.* **2020**, *4* (9), 93801.

(32)    Laws, K. J.; Miracle, D. B.; Ferry, M. A Predictive Structural Model for Bulk Metallic Glasses. *Nat. Commun.* **2015**, *6*. https://doi.org/10.1038/ncomms9123.

(33)    Zeng, S.; Zhao, Y.; Li, G.; Wang, R.; Wang, X.; Ni, J. Atom Table Convolutional Neural Networks for an Accurate Prediction of Compounds Properties. *npj Comput. Mater.* **2019**, *5* (1), 1–7. https://doi.org/10.1038/s41524-019-0223-y.

(34)    Bartel, C. J.; Trewartha, A.; Wang, Q.; Dunn, A.; Jain, A.; Ceder, G. A Critical Examination of Compound Stability Predictions from Machine-Learned Formation Energies. *npj Comput. Mater.* **2020**, *6* (1), 1–11. https://doi.org/10.1038/s41524-020-00362-y.

(35)    Natarajan, A. R.; Van der Ven, A. Machine-Learning the Configurational Energy of Multicomponent Crystalline Solids. *npj Comput. Mater.* **2018**, *4* (1), 1–7. https://doi.org/10.1038/s41524-018-0110-y.

(36)    Kirklin, S.; Saal, J. E.; Meredig, B.; Thompson, A.; Doak, J. W.; Aykol, M.; Rühl, S.; Wolverton, C. The Open Quantum Materials Database (OQMD): Assessing the Accuracy of DFT Formation Energies. *npj Comput. Mater.* **2015**, *1* (October). https://doi.org/10.1038/npjcompumats.2015.10.

(37)   Shapeev, A. V. Moment Tensor Potentials: A Class of Systematically Improvable Interatomic Potentials. *Multiscale Model. Simul.* **2016**, *14* (3), 1153–1173.

(38)   Bartók, A. P.; Payne, M. C.; Kondor, R.; Csányi, G. Gaussian Approximation Potentials: The Accuracy of Quantum Mechanics, without the Electrons. *Phys. Rev. Lett.* **2010**, *104* (13), 136403.

(39)   Fujikake, S.; Deringer, V. L.; Lee, T. H.; Krynski, M.; Elliott, S. R.; Csányi, G. Gaussian Approximation Potential Modeling of Lithium Intercalation in Carbon Nanostructures. *J. Chem. Phys.* **2018**, *148* (24), 241714.

(40)   Xie, T.; Grossman, J. C. Crystal Graph Convolutional Neural Networks for an Accurate and Interpretable Prediction of Material Properties. *Phys. Rev. Lett.* **2018**, *120* (14), 145301.

(41)   Behler, J.; Parrinello, M. Generalized Neural-Network Representation of High-Dimensional Potential-Energy Surfaces. *Phys. Rev. Lett.* **2007**, *98* (14), 146401.

(42)   Zhang, J.; Lei, Y.-K.; Zhang, Z.; Chang, J.; Li, M.; Han, X.; Yang, L.; Yang, Y. I.; Gao, Y. Q. A Perspective on Deep Learning for Molecular Modeling and Simulations. *J. Phys. Chem. A* **2020**, *124* (34), 6745–6763.

(43)   Jiang, D.; Wu, Z.; Hsieh, C. Y.; Chen, G.; Liao, B.; Wang, Z.; Shen, C.; Cao, D.; Wu, J.; Hou, T. Could Graph Neural Networks Learn Better Molecular Representation for Drug Discovery? A Comparison Study of Descriptor-Based and Graph-Based Models. *J. Cheminform.* **2021**, *13* (1), 1–23. https://doi.org/10.1186/s13321-020-00479-8.

(44)   Faber, F. A.; Hutchison, L.; Huang, B.; Gilmer, J.; Schoenholz, S. S.; Dahl, G. E.;

Vinyals, O.; Kearnes, S.; Riley, P. F.; Von Lilienfeld, O. A. Prediction Errors of Molecular Machine Learning Models Lower than Hybrid DFT Error. *J. Chem. Theory Comput.* **2017**, *13* (11), 5255–5264.

(45) Yao, K.; Herr, J. E.; Toth, D. W.; Mckintyre, R.; Parkhill, J. The TensorMol-0.1 Model Chemistry: A Neural Network Augmented with Long-Range Physics. *Chem. Sci.* **2018**, *9* (8), 2261–2269.

(46) Himanen, L.; Jäger, M. O. J.; Morooka, E. V.; Federici Canova, F.; Ranawat, Y. S.; Gao, D. Z.; Rinke, P.; Foster, A. S. DScribe: Library of Descriptors for Machine Learning in Materials Science. *Comput. Phys. Commun.* **2020**, *247*, 106949. https://doi.org/10.1016/j.cpc.2019.106949.

(47) Rupp, M.; Tkatchenko, A.; Müller, K. R.; Von Lilienfeld, O. A. Fast and Accurate Modeling of Molecular Atomization Energies with Machine Learning. *Phys. Rev. Lett.* **2012**, *108* (5), 1–5. https://doi.org/10.1103/PhysRevLett.108.058301.

(48) Faber, F.; Lindmaa, A.; Lilienfeld, O. A. Von; Armiento, R. Crystal Structure Representations for Machine Learning Models of Formation Energies. **2015**. https://doi.org/10.1002/qua.24917.

(49) Behler, J. Atom-Centered Symmetry Functions for Constructing High-Dimensional Neural Network Potentials. *J. Chem. Phys.* **2011**, *134* (7), 74106.

(50) Bartók, A. P.; Kondor, R.; Csányi, G. On Representing Chemical Environments. *Phys. Rev. B* **2013**, *87* (18), 184115.

(51) Pham, T. L.; Kino, H.; Terakura, K.; Miyake, T.; Tsuda, K. Machine Learning Reveals Orbital Interaction in Materials. *Sci. Technol. Adv. Mater.* **2017**, *6996* (November), 1–2. https://doi.org/10.1080/14686996.2017.1378060.

(52)  Jain, A.; Ong, S. P.; Hautier, G.; Chen, W.; Richards, W. D.; Dacek, S.; Cholia, S.; Gunter, D.; Skinner, D.; Ceder, G. Commentary: The Materials Project: A Materials Genome Approach to Accelerating Materials Innovation. *APL Mater.* **2013**, *1* (1), 11002.

(53)  Saal, J. E.; Kirklin, S.; Aykol, M.; Meredig, B.; Wolverton, C. Materials Design and Discovery with High-Throughput Density Functional Theory: The Open Quantum Materials Database (OQMD). *Jom* **2013**, *65* (11), 1501–1509.

(54)  Curtarolo, S.; Setyawan, W.; Hart, G. L. W.; Jahnatek, M.; Chepulskii, R. V; Taylor, R. H.; Wang, S.; Xue, J.; Yang, K.; Levy, O. AFLOW: An Automatic Framework for High-Throughput Materials Discovery. *Comput. Mater. Sci.* **2012**, *58*, 218–226.

(55)  Bergerhoff, G.; Hundt, R.; Sievers, R.; Brown, I. D. The Inorganic Crystal Structure Data Base. *J. Chem. Inf. Comput. Sci.* **1983**, *23* (2), 66–69.

(56)  Awad, M.; Khanna, R. Support Vector Regression. In *Efficient learning machines*; Springer, 2015; pp 67–80.

(57)  Balabin, R. M.; Lomakina, E. I. Support Vector Machine Regression (LS-SVM)—an Alternative to Artificial Neural Networks (ANNs) for the Analysis of Quantum Chemistry Data? *Phys. Chem. Chem. Phys.* **2011**, *13* (24), 11710–11718.

(58)  Kondati Natarajan, S.; Behler, J. Self-Diffusion of Surface Defects at Copper–Water Interfaces. *J. Phys. Chem. C* **2017**, *121* (8), 4368–4383.

(59)  Behler, J.; Parrinello, M. Generalized Neural-Network Representation of High-Dimensional Potential-Energy Surfaces. *Phys. Rev. Lett.* **2007**, *98* (14), 1–4. https://doi.org/10.1103/PhysRevLett.98.146401.

(60)  Behler, J. Four Generations of High-Dimensional Neural Network Potentials. *Chem. Rev.* **2021**, *121* (16), 10037–10072.

(61)    Graser, J.; Kauwe, S. K.; Sparks, T. D. Machine Learning and Energy Minimization Approaches for Crystal Structure Predictions: A Review and New Horizons. *Chem. Mater.* **2018**, *30* (11), 3601–3612.

(62)    Jung, H.; Park, M.; Yoon, Y.-G.; Kim, G.-B.; Joo, S.-K. Amorphous Silicon Anode for Lithium-Ion Rechargeable Batteries. *J. Power Sources* **2003**, *115* (2), 346–351.

(63)    Ohara, S.; Suzuki, J.; Sekine, K.; Takamura, T. A Thin Film Silicon Anode for Li-Ion Batteries Having a Very Large Specific Capacity and Long Cycle Life. *J. Power Sources* **2004**, *136* (2), 303–306.

(64)    Chang, K. C.; Nuhfer, N. T.; Porter, L. M.; Wahab, Q. High-Carbon Concentrations at the Silicon Dioxide-Silicon Carbide Interface Identified by Electron Energy Loss Spectroscopy. *Appl. Phys. Lett.* **2000**, *77* (14), 2186–2188. https://doi.org/10.1063/1.1314293.

(65)    Pujahari, R. M. *Crystalline Silicon Solar Cells*; 2021. https://doi.org/10.1016/B978-0-12-823710-6.00004-2.

(66)    Guha, S.; Yang, J.; Nath, P.; Hack, M. Enhancement of Open Circuit Voltage in High Efficiency Amorphous Silicon Alloy Solar Cells. *Appl. Phys. Lett.* **1986**, *49* (4), 218–219. https://doi.org/10.1063/1.97176.

(67)    Yanxon, H.; Zagaceta, D.; Wood, B. C.; Zhu, Q. Neural Network Potential from Bispectrum Components: A Case Study on Crystalline Silicon. *J. Chem. Phys.* **2020**, *153* (5), 54118.

(68)    Comin, M.; Lewis, L. J. Deep-Learning Approach to the Structure of Amorphous Silicon. *Phys. Rev. B* **2019**, *100* (9), 94107.

(69)    Jain, A.; Hautier, G.; Moore, C. J.; Ping Ong, S.; Fischer, C. C.; Mueller, T.; Persson, K.

A.; Ceder, G. A High-Throughput Infrastructure for Density Functional Theory Calculations. *Comput. Mater. Sci.* **2011**, *50* (8), 2295–2310. https://doi.org/10.1016/j.commatsci.2011.02.023.

(70) Peterson, G. G. C.; Brgoch, J. Materials Discovery through Machine Learning Formation Energy. *JPhys Energy* **2021**, *3* (2). https://doi.org/10.1088/2515-7655/abe425.

(71) Chen, W. C.; Vohra, Y. K.; Chen, C. C. Discovering Superhard B-N-O Compounds by Iterative Machine Learning and Evolutionary Structure Predictions. *ACS Omega* **2022**, *7* (24), 21035–21042. https://doi.org/10.1021/acsomega.2c01818.

(72) Ward, L.; Dunn, A.; Faghaninia, A.; Zimmermann, N. E. R.; Bajaj, S.; Wang, Q.; Montoya, J.; Chen, J.; Bystrom, K.; Dylla, M.; Chard, K.; Asta, M.; Persson, K. A.; Snyder, G. J.; Foster, I.; Jain, A. Matminer: An Open Source Toolkit for Materials Data Mining. *Comput. Mater. Sci.* **2018**, *152* (April), 60–69. https://doi.org/10.1016/j.commatsci.2018.05.018.

(73) Ong, S. P.; Richards, W. D.; Jain, A.; Hautier, G.; Kocher, M.; Cholia, S.; Gunter, D.; Chevrier, V. L.; Persson, K. A.; Ceder, G. Python Materials Genomics (Pymatgen): A Robust, Open-Source Python Library for Materials Analysis. *Comput. Mater. Sci.* **2013**, *68*, 314–319.

(74) Awad, M.; Khanna, R. Support Vector Regression. *Effic. Learn. Mach.* **2015**, No. November 2007, 67–80. https://doi.org/10.1007/978-1-4302-5990-9_4.

(75) Pedregosa, F.; Varoquaux, G.; Gramfort, A.; Michel, V.; Thirion, B.; Grisel, O.; Blondel, M.; Prettenhofer, P.; Weiss, R.; Dubourg, V.; Vanderplas, J.; Passos, A.; Cournapeau, D.; Brucher, M.; Perrot, M.; Duchesnay, É. Scikit-Learn: Machine Learning in Python. *J. Mach. Learn. Res.* **2011**, *12*, 2825–2830.

(76)   Waskom, M. L. Seaborn: Statistical Data Visualization, J. Open Source Softw., 6, 3021. 2021.

(77)   Aggarwal, V.; Gupta, V.; Singh, P.; Sharma, K.; Sharma, N. Detection of Spatial Outlier by Using Improved Z-Score Test. In *2019 3rd International Conference on Trends in Electronics and Informatics (ICOEI)*; IEEE, 2019; pp 788–790.

(78)   Kim, S.; Noh, J.; Gu, G. H.; Aspuru-Guzik, A.; Jung, Y. Generative Adversarial Networks for Crystal Structure Prediction. *ACS Cent. Sci.* **2020**, *6* (8), 1412–1420.

(79)   Cao, Z.; Dan, Y.; Xiong, Z.; Niu, C.; Li, X.; Qian, S. Convolutional Neural Networks for Crystal Material Property Prediction Using Hybrid Orbital-Field Matrix and Magpie Descriptors. **2019**, 1–15. https://doi.org/10.3390/cryst9040191.